\documentclass[twocolumn,prl,superscriptaddress,nobib]{revtex4}
\usepackage{color}
\definecolor{red}{rgb}{0.75,0,0}
\definecolor{blue}{rgb}{0,0,0.75}
\definecolor{green}{rgb}{0,0.5,0}
\usepackage[pdftex, pdfborder={0 0 0}, colorlinks=true, linkcolor=red, urlcolor=blue]{hyperref}
\usepackage{amsmath}
\usepackage{amssymb}
\usepackage{bm}
\usepackage{graphicx}
\usepackage{verbatim}
\usepackage{changes}
\usepackage{lipsum}% <- For dummy text
\definechangesauthor[name={Per cusse}, color=orange]{per}
\setremarkmarkup{(#2)}

\def\be{\begin{equation}}
\def\ee{\end{equation}}
\def\bea{\begin{eqnarray}}
\def\eea{\end{eqnarray}}

\def\besub{\begin{subequations}}
\def\eesub{\end{subequations}}

\def\bwd{\begin{widetext}}
\def\ewd{\end{widetext}}

\newcommand{\bsf}[1]{\textsf{\textbf{#1}}}

\begin{document}
\title{Stability from activity}
\author{Ananyo Maitra}
\email{ananyo.maitra@u-psud.fr}
\affiliation{LPTMS, CNRS, Univ. Paris-Sud, Universit\'e Paris-Saclay, 91405 Orsay, France}
\author{Pragya Srivastava}
\affiliation{The Francis Crick Institute, Lincoln’s Inn Fields Laboratory, 44 Lincoln’s Inn Fields,
London WC2A 3LY}
\author{M. Cristina Marchetti}
\affiliation{Physics Department and Syracuse Soft Matter Program, Syracuse University, Syracuse,
NY 13244, USA}
\author{Juho Lintuvuori}
\affiliation{Univ. Bordeaux, CNRS, LOMA, UMR 5798, F-33405 Talence, France }
\author{Sriram Ramaswamy}
\affiliation{Indian Institute of Science, Bangalore 560012, India}
\author{Martin Lenz}
\email{martin.lenz@u-psud.fr}
\affiliation{LPTMS, CNRS, Univ. Paris-Sud, Universit\'e Paris-Saclay, 91405 Orsay, France}

\begin{abstract}
Suspensions of actively driven anisotropic objects exhibit distinctively nonequilibrium behaviors, and current theories predict that they are incapable of sustaining orientational order at high activity. By contrast, here we show that {nematic} suspensions on a substrate can display order at arbitrarily high activity due to a previously unreported, potentially stabilizing active force. The resulting {nonequilibrium ordered phase displays} robust giant number fluctuations that cannot be suppressed even by an incompressible solvent. Our results apply to virtually all experimental assays used to investigate the active nematic ordering of self-propelled colloids, bacterial suspensions and the cytoskeleton, and have testable implications in interpreting their nonequilibrium behaviors.
\end{abstract}

\maketitle

%\remark{You need to reorder the references throughout. (I'll never understand why you don't use bibtex); you also need to complete the references to the PRL format (not all of ther)}

Living systems convert chemical energy into motion, thus violating {detailed balance} at the microscopic scale. Macroscopically, these violations result in stresses and currents responsible for intracellular flows leading to cellular motion~\cite{Kruse}, collective cell migration during embryonic development \cite{Silberzan} and the flocking of birds~\cite{Cavagna}. Similar nonequilibrium currents arise in non-living systems such as chemotactic colloids~\cite{Suropriya} and vibrated granular rods~\cite{Vijay}. 
These systems are often described by active hydrodynamic theories, a class of continuum descriptions derived from equilibrium  theories of liquid crystals but supplemented with extra ``active" forces arising from microscopic driving~\cite{SR_lyon, RMP}. These theories are tools of choice to study specifically nonequilibrium features in the collective behaviors of fluid suspensions of anisotropic active units such as cytoskeletal filaments \cite{Dogic, Weber} or bacteria \cite{Goldstein}.

{A central issue in active hydrodynamics is to determine the effects of activity on the dynamic stability, and the robustness against fluctuations, of various types of orientational and translational order.} Previous studies have shown that{,} due to {the interplay of active stress and solvent flow}, nematic order in incompressible active suspensions is always unstable beyond a critical value of activity \cite{RMP, Simha_Ramaswamy, Voit1, Marenduzzo}. This instability threshold vanishes {in the limit of infinite system size}, implying that, unlike their equilibrium counterparts, these systems are generically unstable. In  two-dimensional experimental realizations this instability can, however, be suppressed by the friction of the fluid against a substrate. Nevertheless, current theories predict that even under these conditions instability always occurs at high enough activity \cite{Ananyo_nucrot}, which may be experimentally realized through an increase of the density of myosin motors or bacteria, or of the amount of fuel available to them.

Another distinctive feature of active systems is the statistics of their density fluctuations. In equilibrium systems {away from critical points and with finite-range interactions,} a region of space containing $N$ particle on average will undergo fluctuations of this number of order $\sqrt{N}$ whether or not it is embedded in an incompressible solvent. In contrast, active hydrodynamic theories of systems without incompressibility display fluctuations of order greater than $\sqrt{N}$ due to active mass currents arising from orientation fluctuations \cite{RMP, Ramaswamy_Simha_Toner, suraj}. While these so-called giant number fluctuations have clearly been observed in solvent-less simulations, little is known about their form in the presence of an incompressible solvent \cite{Shradha_thesis}, and their observations in biological experiment has been difficult and controversial \cite{Chate_new}.

%To help interpret the rich dynamical behaviors of current, typically quasi-two-dimensional experiments on active systems [8, 9, 16, 17] here we theoretically study the ordering and fluctuations of a apolar active fluid in contact with a substrate. Here are our main results. (i) Re-examining the origins of active hydrodynamic theories in two dimensions we find that contact with a substrate allows a new active force with a distinct angular symmetry. This force does not conserve angular momentum, yet it exists even in achiral systems. (ii) Contrary to common wisdom, increased activity – in the presence of this new term – can lead to an enhanced stabilisation of nematic ordering. (iii) Giant number fluctuations in the active nematic phase are robust to the introduction of incompressible solvent as well as our new active force. (iv) The new active force emerges naturally from an accepted model of three-dimensional active fluid under vertical confinement, and is shown to be the dominant active force in a renormalized theory in the presence of noise. (v) Our results offer plausible explanations for the persistence of order at high activity, as well as the systematics of instabilities, in bacterial, living liquid crystal and cytoskeletal systems.

To help interpret the rich dynamical behaviour of current, typically quasi-two-dimensional experiments on active systems \cite{Chate_new, Dogic,Goldstein, Zhou} here we study theoretically the ordering and fluctuations of an apolar active fluid in contact with a substrate.
By re-examining the foundation of active hydrodynamic theories in two dimensions, we first find that the contact with a substrate allows a new active force with a distinct angular symmetry. This force does not conserve angular momentum, yet it exists even in achiral systems. Contrary to common wisdom, increased activity in the presence of this new term can lead to a {\it stabilisation} of nematic ordering. 
Next, we show that giant number fluctuations in the active nematic phase are robust to the introduction of incompressible solvent as well as of our new active force.
The new active force emerges naturally from an accepted model of three-dimensional active fluid under vertical confinement, and we argue that it is the dominant active force in a renormalized theory in the presence of noise. Our results offer plausible explanations for the persistence of order at high activity, as well as the systematics of instabilities, in bacterial, living liquid crystal and cytoskeletal systems.

We consider a nematically ordered two-dimensional active {suspension} described by the local orientation $\theta(\mathbf{x},t)$ and concentration $c(\mathbf{x},t)$ of its active particles, and centre-of-mass velocity $\mathbf{v}(\mathbf{x},t)$. Considering small deviations from a homogeneous state aligned along $\mathbf{\hat{x}}$, we write the {linear} dynamical equations compatible with the symmetries of the system in the long-wavelength (hydrodynamic) limit. The {dynamical equation for the} angle field {is}
\begin{equation}
\label{orient}
\dot{\theta}=\frac{1-\lambda}{2}\partial_xv_y-\frac{1+\lambda}{2}\partial_yv_x-\Gamma_\theta\frac{\delta  \mathcal{H}}{\delta\theta},
\end{equation}
where $|\lambda|>1$ describes particles with a tendency to align under a shear flow (\emph{e.g.}, cytoskeletal filaments) while $|\lambda|<1$ denotes flow tumbling (as in bacteria). Here we use  the simplified one-Frank-constant free-energy functional
\begin{equation}\label{eq:freeenergy}
\mathcal{H}=\int d^2{\bf r} \left[\frac{K}{2}(\nabla \theta)^2+g(c)\right],
\end{equation}
where $K>0$ characterizes the tendency of the particles to align and $g(c)$ is an arbitrary function of the concentration.

\begin{figure}
\centering
\includegraphics[width=\columnwidth]{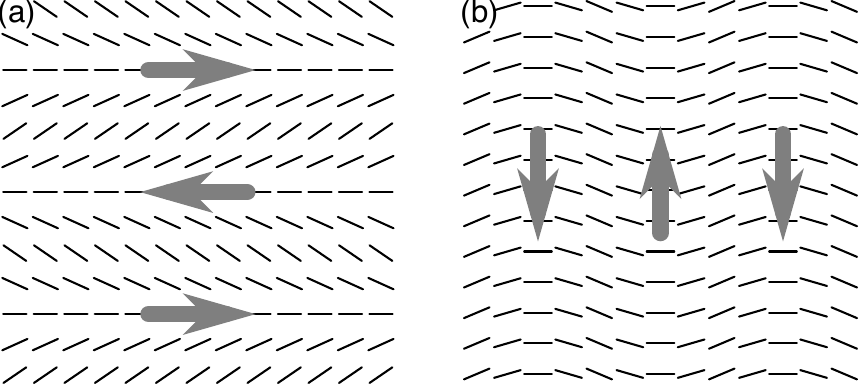}
\caption{\label{fig:activeterms}
The two components of the active force $\mathbf{f}^a$ determine the stability of the active nematic with respect to splay and bend. (a)~A splay perturbation $\partial_y\theta\neq 0$ (black segments) induces an active force $f_x^a=a\partial_y\theta$ which stabilizes a flow-tumbling system if $a<0$ (as represented by the arrows). (b)~A bend perturbation $\partial_x\theta\neq 0$ produces an active force $f_y^a=b\partial_x\theta$ which stabilizes a flow-tumbling system if $b>0$. Our new active force opens the possibility for both forces to be stabilizing at the same time. 
}
\end{figure}

Flow is driven by forces internal to the fluid, and the presence of the substrate dictates a Darcy dynamics:
\begin{equation}\label{eq:Darcy}
\Gamma \mathbf{v}=-\nabla\Pi+\mathbf{f}^p+\mathbf{f}^a,
\end{equation}
where $\Gamma$ can be viewed as the friction coefficient against the substrate. The {pressure $\Pi$ serves as a Lagrange multiplier enforcing the incompressibility condition $\nabla\cdot\mathbf{v}=0$ for the suspension as a whole, while still permitting fluctuations in the concentration of suspended particles.} We will not consider the case of a compressible medium, which adds no physics of interest. Onsager symmetry and Eq.~(\ref{orient}) yield the density of passive (equilibrium) forces
\begin{equation}
{\bf f}^p=-\frac{1+\lambda}{2}\partial_y\left(\frac{\delta  \mathcal{H}}{\delta\theta}\right)\mathbf{\hat{x}}+\frac{1-\lambda}{2}\partial_x\left(\frac{\delta  \mathcal{H}}{\delta\theta}\right)\mathbf{\hat{y}}.
\end{equation}
Beyond these standard equilibrium terms, the active force density $\mathbf{f}^a$ depends on $\theta$ only through its gradient due to rotational invariance. Combining this with the $({x,} y,\theta)\rightarrow ({x,} -y,-\theta)$ reflection invariance of our achiral system dictates that to lowest order in gradients
\begin{subequations}\label{eq:activeforce}
\begin{eqnarray}
f_x^a&=&-(\zeta_1\Delta\mu+\zeta_2\Delta\mu)\partial_y\theta\\
f_y^a&=&-(\zeta_1\Delta\mu-\zeta_2\Delta\mu)\partial_x\theta,
\end{eqnarray}
\end{subequations}
where $\zeta_1$ and $\zeta_2$ are two independent, \emph{a priori} unknown phenomenological constants, and $\Delta\mu$ denotes the strength of the overall activity in the system, \emph{e.g.}, the chemical potential difference between the cellular fuel ATP and its hydrolysis products. We interpret the two components of $\mathbf{f}^a$ in Fig.~\ref{fig:activeterms}, with $f_x^a$ inducing a horizontal fluid flow in a splayed nematic while $f_y^a$ drives a vertical fluid flow in a bent nematic. An active force {depending} on gradients of $c$ is also allowed in $\mathbf{f}^a$, but does not significantly modify our discussion~\cite{supp}. While the active force proportional to $\zeta_1$ is standard in active fluid theories, the $\zeta_2$ force is presented here for the first time.

This new force can be understood in simple terms by introducing the nematic director ${\bf n}=(\cos\theta,\,\sin\theta)$. In momentum-conserving systems, the active force can only be the divergence of a symmetric stress, namely $\mathbf{f}^a=\zeta_1\,\Delta\mu\nabla\cdot({\bf nn})=\zeta_1\Delta\mu\left[{\bf n}(\nabla\cdot{\bf n})+{\bf n}\cdot\nabla {\bf n}\right]$. Our new force involves an exchange of angular momentum with the substrate, and introducing it thus allows 
an active force with different prefactors for the terms ${\bf n}(\nabla\cdot{\bf n})$ and ${\bf n}\cdot\nabla {\bf n}$.

The evolution of the concentration $c$ is governed by a conservation equation $\partial_t c = -\nabla\cdot(\mathbf{J}^p+\mathbf{J}^a)$, where the passive particle current reads $\mathbf{J}^p=-\Gamma_c \nabla{\delta{\cal H}}/{\delta c}$. As {with} $\mathbf{f}^a$, the active current $\mathbf{J}^a$ comprises {of} two distinct {$\theta$-dependent} active terms \emph{a priori}, but the {term analogous to $\zeta_2$, being a curl, drops out of the conservation equation}, yielding
\begin{equation}
\label{conc}
\partial_t c=\Gamma_c\nabla^2\frac{\delta \mathcal{H}}{\delta c}+\zeta_c\Delta\mu\partial_x\partial_y\theta,
\end{equation}
where the $\zeta_c\Delta\mu$ active term couples orientation fluctuations with concentration fluctuations, and is featured in standard theories of active nematics~\cite{Ramaswamy_Simha_Toner, RMP}.

\begin{figure}
 \includegraphics[width=\columnwidth]{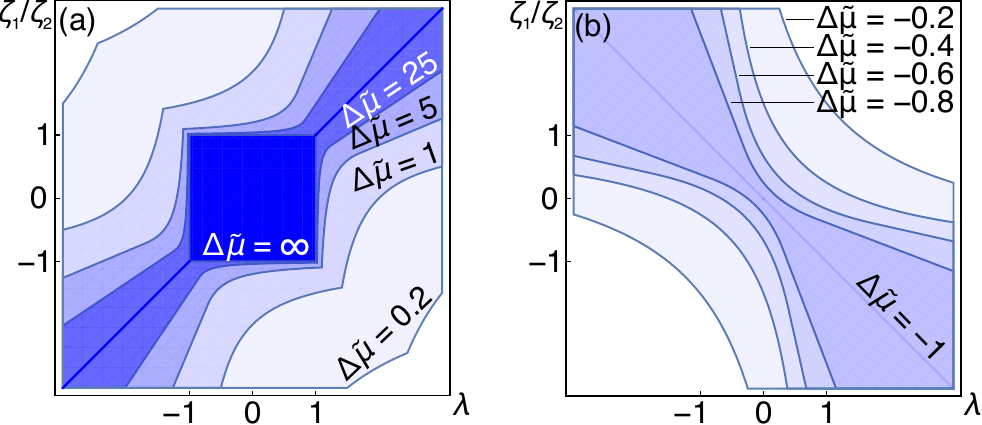}
\caption{
Regions of stability of the ordered phase as given by the sign of $D(\phi)$ in Eq.~(\ref{diffusivity}) as a function of the flow-alignment parameter $\lambda$, the ratio $\zeta_1/\zeta_2$ of the old and new active forces and the overall magnitude of activity relative to passive friction $\Delta\tilde{\mu}=\zeta_2\Delta\mu/{2K\Gamma\Gamma_\theta}$. (a)~For $\Delta\tilde{\mu}>0$, the region of linear stability of the ordered phase (shades of blue) shrinks with increasing activity, yet the central dark blue square is stable for arbitrary high activity. (b)~For $\Delta\tilde{\mu}<0$, stability is abolished for large enough activity, namely $\Delta\tilde{\mu}<-1$.
}
\label{fig:linearStability}
\end{figure}

The new active term $\zeta_2\Delta\mu$ has dramatic consequences for the linear stability of the active fluid. 
Consider the evolution {of} a small perturbation $\theta_\mathbf{q}e^{i\mathbf{q}\cdot\mathbf{x}}$ with a wave vector $\mathbf{q}=q(\cos\phi\mathbf{\hat{x}}+\sin\phi\mathbf{\hat{y}})$. Combining Eqs.~(\ref{orient} {-} \ref{conc}) and eliminating pressure by projecting on the direction perpendicular to $\mathbf{q}$, we find that the dynamics is diffusive {to leading order in $q$}: $\partial_t \theta_\mathbf{q}=-D(\phi)q^2\theta_\mathbf{q}$, where the direction-dependent {orientational} diffusivity is given by:
\begin{equation}
\label{diffusivity}
D(\phi)=\Gamma_\theta K+\frac{\Delta\mu}{2\Gamma}(1-\lambda\cos 2\phi)(-\zeta_1\cos 2\phi+\zeta_2).
\end{equation}
{The system is thus linearly stable at small $q$} if and only if $D(\phi)$ is positive for all values of $\phi$, {i.e., if the second term on the right in \eqref{diffusivity} is not so large as to overcome the stabilizing effect of director relaxation through} $\Gamma_\theta K$. We {now focus on this second term, which dominates for high activity, \emph{i.e.}, large $\Delta\mu$}. {The case of flow-tumbling ($|\lambda|<1$) systems, where {$(1-\lambda\cos 2\phi)$} is always positive, is easy to interpret as it is controlled by} the active force $\mathbf{f}^a$. As shown in Fig.~\ref{fig:activeterms}, splay can only be stabilized by a force $f_x^a$ that depends on $\partial_y\theta$ through a negative coefficient, {whereas} bend stabilization requires $f_y^a$ to depend on $\partial_x\theta$ through a positive coefficient. As previous studies implicitly assume $\zeta_2=0$, Eq.~(\ref{eq:activeforce}) clearly shows that they impose the equality of these two coefficients, implying a destabilization of either bend or splay depending on its sign. The inclusion of the new active force ($\zeta_2\neq 0$) now offers the possibility for these coefficients to have opposite signs. {For $\zeta_2>|\zeta_1|$ this implies that the stability of the system, i.e., a positive relaxation rate,} increases with increasing activity, as shown in Fig.~\ref{fig:linearStability}. By contrast, {both} flow-aligning ($|\lambda|>1$) and $\zeta_2<|\zeta_
1|$ systems remain generically unstable at high activity.

Having opened up the possibility of a stable homogeneous nematic at high activity, we now examine the nature of concentration fluctuations arising in such a state when non-conserving and conserving noise sources are added to Eqs.~\eqref{orient} and \eqref{conc} respectively. We find that the giant number fluctuations persist despite the long-range effects associated with the incompressible velocity field. This result, which is in clear contrast to the case of incompressible active polar systems \cite{Ananyo_polar, Bartolo}, can be seen without a detailed calculation by examining the structure of Eqs.~\eqref{conc} and \eqref{diffusivity}. 
 Indeed, compared to compressible nematic systems, incompressibility only introduces a non-singular anisotropy in the orientational relaxation rate \eqref{diffusivity}, without modifying the scaling with wavenumber. Since giant number fluctuations rely solely on the wavenumber scaling of the orientational relaxation rate and not on its anisotropy \cite{Ramaswamy_Simha_Toner}, they should be present in our system as well. Specifically, simple power-counting within a linearized treatment demonstrates that, due to the active concentration current, concentration fluctuations $\delta c=c-\langle c\rangle$ scale with $q$ in the same way as orientation fluctuations. Therefore, the static structure factor  $\mathcal{S}_{\bf q} \equiv \int_{\bf r} \langle \delta c({\bf 0}) \delta c({\bf r})\rangle \exp(-i {\bf q} \cdot {\bf r}) / \langle c\rangle$ of concentration fluctuations scales as $1/q^2$. Equivalently, in a region containing on average $N$ particles, the standard deviation in the number scales linearly with $N$ in two dimensions. Only the form of the predicted anisotropy of $\mathcal{S}_{\bf q}$ is modified \cite{supp} with respect to that in \cite{Ramaswamy_Simha_Toner}, thanks to the factor $D(\phi)$ in Eq.~\eqref{diffusivity}.
Our conclusions remain valid even upon inclusion of more general symmetry-allowed active and passive terms \cite{supp}.

To {further} elucidate the physical origin of our new active force, we consider the dynamics of a three-dimensional active {suspension} of lateral dimension $L$ confined over a length scale $h\ll L$ in the $z$-direction and project it {onto a}  two-dimensional $xy${-plane}. This projection involves a standard lubrication approximation \cite{Stone} supplemented by a mean-field treatment whereby the $z$-average of a product of two fields is {approximated by} the product of their individual $z$-averages, and is detailed in the supplemental {material \cite{supp}}. Denoting three-dimensional fields and operators with a bar, we describe our fluid in terms of its three-dimensional nematic order parameter $\bar{\bsf{Q}}(\bar{\mathbf{r}},t)$, velocity $\bar{\mathbf{v}}(\bar{\mathbf{r}},t)$ and density $\bar{c}(\bar{\mathbf{r}},t)$. Its evolution follows well-established active fluid equations, namely {three-dimensional} evolution equations for the nematic order parameter and concentration physically similar to Eqs.~(\ref{orient}) and (\ref{conc}), as well as the force balance condition for an overdamped viscous fluid in presence of an active stress tensor
\begin{equation}\label{eq:3DActiveStress}
\bar{\boldsymbol{\sigma}}=
	\bar{\zeta}_0\Delta\mu\,\bar{\bsf{I}}
	-\bar{\zeta}_1\Delta\mu\,\bar{\bsf{Q}}
	-\bar{\zeta}_2\Delta\mu\,\bar{\nabla}\left[\bar{\bsf{Q}}\cdot\left(\bar{\nabla}\cdot\bar{\bsf{Q}}\right)\right],
\end{equation}
where $\bar{\bsf{I}}$ denotes the unit tensor.
While the two first terms in the right-hand-side of Eq.~(\ref{eq:3DActiveStress}) are standard, the last term {would be disregarded in a gradient-expansion treatment of an unconfined fluid. Counterintuitively, however, it is this term, in a thickness-averaged description, which leads to the second activity constant $\zeta_2$, thanks to the dominance of gradients $\sim 1/h$ along the $z$ direction.}   

Averaging our three-dimensional equations in the $z$ direction and projecting them onto the $xy$ plane, we obtain coupled equations of motion for the thickness-averaged planar velocity $\mathbf{v}(\mathbf{x},t)$ and two-dimensional nematic tensor 
	\begin{equation}
\label{apolar_OP}
\bsf{Q}(\mathbf{x},t)=\frac{S}{2}\begin{pmatrix}\cos 2\theta &\sin 2\theta\\ \sin 2\theta &{-}\cos2\theta\end{pmatrix},
\end{equation}
where $S$ denotes the magnitude of the order parameter. These equations read
\begin{subequations}\label{eq:2Dequations}
\begin{eqnarray}
\partial_t \bsf{Q} &=&
-{\bf v} \cdot \nabla {{\bsf{Q}}}
+\boldsymbol{{\omega}}\cdot {{\bsf{Q}}}
-{{\bsf{Q}}}\cdot{\boldsymbol{{\omega}}}-\lambda\bsf{U}
-\Gamma_\theta \bsf{H}
\label{Q2d},
\\
\Gamma {\bf v}&=&-\nabla {\Pi}
-\lambda\nabla\cdot\bsf{H}
-2\nabla\cdot(\bsf{QH})^A,\nonumber
\\
&&-{\zeta}_1\Delta\mu\nabla\cdot
\bsf{Q}-2\zeta_2\Delta\mu\,\bsf{Q}\cdot\left(\nabla\cdot
\bsf{Q}\right),
\label{aver}
\end{eqnarray}
\end{subequations}
where $\zeta_1=\bar{\zeta}_1$ and $\zeta_2=9\bar{\zeta}_2/h^2$ and the dynamics of the concentration field is described by Eq.~\eqref{conc}. These equations reduce to Eqs.~\eqref{orient}-\eqref{eq:activeforce} deep in the ordered phase where {$S$ relaxes within a microscopic time} to its steady-state value, which we set to $1$ without loss of generality. In Eq.~\eqref{eq:2Dequations} $\bsf{H}$ is the molecular field conjugate to ${\bsf Q}$, the superscript $A$ denotes the antisymmetric part of a tensor, and $\boldsymbol{{\omega}}$ and $\bsf{U}$ are respectively the antisymmetric and symmetric parts of the tensor $\nabla\mathbf{v}$. The pressure $\Pi$ imposes incompressibility as in Eq.~\eqref{eq:Darcy}. Eqs.~\eqref{eq:2Dequations} demonstrate that the new force $\propto \zeta_2$ previously introduced through general symmetry arguments is a natural emergent feature of a confined three-dimensional active dynamics. Note that while we have included only one term of order $\bar{\nabla}^2$ in Eq.~(\ref{eq:3DActiveStress}) for clarity, other ${\cal O}(\bar{\nabla}^{n\geqslant2})$ terms also contribute {both} to this new active force {and the old one}, {upon thickness-averaging}, although they do not introduce any qualitatively new term. Dimensionally, each additional factor of $\nabla$ in the three-dimensional theory must be accompanied by a corresponding factor of $\ell$, a length scale which should normally be of order the size of the suspended particles. We thus expect $\zeta_2/\zeta_1 \sim (\ell/h)^2$. In closely confined suspensions with $h \sim \ell$, we thus expect $\zeta_2$ and $\zeta_1$ to be comparable in magnitude.

Beyond these microscopic considerations, we predict that the new $\zeta_2$ force will dominate {over} the old $\zeta_1$ force in a renormalized theory in the presence of noise. Indeed, according to Eqs.~\eqref{apolar_OP} and \eqref{aver} the latter reads $\nabla\cdot{\bsf Q}=\cos 2\theta(\partial_y\theta\hat{x}+\partial_x\theta\hat{y})$ in the ordered phase, while the new force takes a form $2{\bsf Q}\cdot(\nabla\cdot{\bsf Q})=\partial_y\theta\hat{x}-\partial_x\theta\hat{y}$ that does not involve the anisotropic factor $\cos 2\theta$. Since active nematics only have quasi-long-range-order, all anisotropic terms average to zero at large scales due to rotation invariance \cite{suraj}, implying that $\langle \cos 2\theta\rangle$, decays as a power of system-size with a typically small exponent. Therefore, for large systems, the isotropic $\zeta_2$ active force does not vanish with diverging system size {\cite{supp}} and thus dominates over the $\zeta_1$ force.

In addition to its role in nematic and polar systems, the new active force introduced here is {the} key to characterising activity in higher-symmetry active systems, including tetractic \cite{Vijay} and hexatic \cite{Chaikin, Aparna1} phases. {In} these systems symmetry imposes that $\zeta_1$, $\zeta_c$, and $\lambda$ in \eqref{orient}-\eqref{conc} all vanish, implying that our $\zeta_2$ term is the only possible source of active instabilities. {It arises through an antisymmetric piece of the active stress, proportional in two dimensions to $\theta \boldsymbol{\epsilon}$, where the pseudoscalar angle field $\theta$ is the broken-symmetry mode and $\boldsymbol{\epsilon}$ is the two-dimensional Levi-Civita tensor.} While this does not give rise to a generic instability, these active $p$-atics on substrates are nevertheless unstable for $\zeta_2\Delta\mu<-\Gamma_\theta K$ irrespective of the presence of an incompressible solvent. {In addition, the aforementioned pure-curl character of the $\zeta_2$ term means that they do not contribute to mass currents. Active tetratics and hexatics thus have normal (non-giant) number fluctuations.}

{We close with the implications of our ideas for several current experiments on biological active matter. Recent experiments ~\cite{Chate_new} find highly ordered apolar nematic phases in confined suspensions of \emph{E. coli} bacteria. The absence of bacterial turbulence in these systems would have been a a puzzle in a treatment with a single activity parameter: the experiments use a non-tumbling mutant, \emph{i.e.}, a highly reduced $\Gamma_\theta$ in \eqref{diffusivity}, which should favour instability. The resolution could well lie in our mechanism involving a second activity parameter with a possibly stabilizing effect.}

{Our linear stability analysis also helps heuristically understand pattern formation in bacterial and cytoskeletal active fluids~\cite{Giomi, Goldstein,Dogic}. The anisotropy of $D(\phi)$ in Eq.~\eqref{diffusivity} implies that the ordered phase is destabilised when it first becomes negative for any angle $\phi_u$. In this case, bands with the normal vector oriented along $\phi_u$ are likely to form whose length scale can be obtained  simply by extending, to fourth order in the wavenumber $q$, the two-dimensional mode analysis, that led to Eq.~(\ref{diffusivity}). The resulting dynamics reads $\partial_t\delta\theta_\mathbf{q}=-[D(\phi)q^2+K_r(\phi)q^4]\delta\theta_\mathbf{q}$, where the stabilizing coefficient $K_r(\phi)=(K/4\Gamma)(1-\lambda\cos 2\phi)^2$ arises simply from Frank elasticity, and thus accounts for patterns with size $\approx\sqrt{K_r(\phi_u)/|D(\phi_u)|}$ without resorting to previously invoked \emph{ad hoc} high-order gradient expansions~\cite{Goldstein,Pragya}. Moreover, when the effective $2D$ dynamics is that of a confined $3D$ fluid as discussed above, $\Gamma \propto 1/h^2$ so that $K_r$ dominates {over} such \emph{ad hoc} terms by a factor ${\cal O}(h/\ell)^2$, which can be large depending on the {scale} of the confinement. }

Beyond these applications, this order-$q^4$ theory can be extended to explain ordering and pattern formation in novel ``living liquid crystals", namely passive nematic liquid crystals with well-characterized physical properties perfused with small quantities of bacteria that offer unprecedented opportunities {for } quantitative tests of active matter theories. Our proposed approach relies on two coupled angle fields for the local alignment of the passive liquid crystalline particles and active bacterial ones, respectively, {as does the model of \cite{Zhou}}. {However, it } accounts for activity without the need for previously introduced \emph{ad hoc} Onsager symmetry breaking orientational couplings between these two angular fields~\cite{Zhou}, and leads to a stability criterion identical to that of Eq.~\eqref{diffusivity}. As the quantity analogous to $\lambda$ can be quantitatively tuned by modifying the passive properties of the liquid crystal, our stability prediction is directly testable, and we predict the appearance of patterns {with } a similarly tunable length scale~\cite{supp}.

{Finally, our approach suggests an explanation for the recent observation \cite{Dogic2} that the transition to activity-driven turbulence in extensile microtubule-kinesin systems confined in a channel is controlled by the aspect ratio of the transverse cross-section of this channel.
Averaging the dynamics over the $z$ direction in a channel with a rectangular $L_y\times h$ cross-section, we look at the longest wavelength splay fluctuations in the $y$ direction, while noting that extensile particles imply $\bar{\zeta}_1>0$, which stabilises splay fluctuations. Assuming a destabilizing $\bar{\zeta}_2<0$, we predict that the system should become unstable as $h$ is decreased and the ratio $\zeta_2/\zeta_1\propto1/h^2$ subsequently increases. 
%As in Eq.~\eqref{diffusivity}, the stability of the ordered phase to the dominant fluctuations along $y$ involves a competition between the passive friction $\Gamma_\theta K/L^2$ and an active term $(\Delta\mu/\eta|\bar{\zeta}_2|)(1+\lambda)[(h/L_y)^2|\bar{\zeta}_1/\bar{\zeta}_2|-1/L_y^2]$, where $\eta$ is the suspension viscosity \remark{Why the absolute values and no angular dependence in the active term?}. Assuming $\bar{\zeta}_2<0$, the active term changes sign upon decreasing the aspect ration $h/L_y$ while holding $L_y$ fixed, leading to instability and eventually to turbulence.
We therefore predict that our new active force plays a destabilizing role in microtubule-kinesin systems.}

\begin{acknowledgments}
This work was supported by Marie Curie Integration Grant PCIG12-GA-2012-334053, ``Investissements d'Avenir'' LabEx PALM (ANR-10-LABX-0039-PALM), ANR grant ANR-15-CE13-0004-03 and ERC Starting Grant 677532. ML's group belongs to the CNRS consortium CellTiss. {SR acknowledges support from a J. C. Bose National Fellowship of the SERB, India and from the Tata Education and Development Trust. MCM was suppoerted by the US National Science Foundation awards NSF-DMR-1609208 and NSF-DGE-1068780 and by the Syracuse Soft Matter Programme. AM, PS, SR and MCM also acknowledge the  support  of  the  Kavli  Institute  for  Theoretical  Physics  under  Grant  No.   NSF  PHY11-25915. JL was supported by IdEx Bordeaux Junior Chair.}
\end{acknowledgments}

\end{document}